\documentstyle[12pt]{article}
\newlength{\aivwidth}   
\newlength{\tmpwidth}   

\newcommand{\lag}{{\cal L}}
\newcommand{\ham}{{\cal H}}
\newcommand{\lib}{\bar{\lag}_I}
\newcommand{\ep}{\epsilon}
\newcommand{\df}{\delta^4(0)}
\newcommand{\Df}{{\cal D}}
\newcommand{\dx}{d^4x\,}
\newcommand{\vp}{\varphi}
\newcommand{\bp}{{\bar{\psi}}}
\newcommand{\vc}{\vp^\dagger}
\newcommand{\vci}{{\vp^\dagger}}
\newcommand{\pa}{\partial}
\newcommand{\tb}{\bar{t}}
\renewcommand{\S}{\scriptstyle}
\newcommand{\subs}{{\begin{array}{l}
     \S F_{i0}^a\to\pi_i^a\\\S D_0\vc_a
           \to\pi_\vp^a\\\S D_0\vp_a\to\pi_\vci^a\end{array}}}
\newcommand{\ssubs}{{\begin{array}{l}
     \S \pi_i^a\to F_{i0}^a\\\S \pi_\vp^a\to
            D_0\vc_a\\\S \pi_\vci^a\to D_0\vp_a\end{array}}}
\newcommand{\Det}{\, {\rm Det}\,}
\newcommand{\vx}{{\bf x}}
\newcommand{\vy}{{\bf y}}
\newcommand{\dzero}{\delta(x^0-y^0)}
\newcommand{\X}{{\rm X}}

\title{Equivalence of Hamiltonian and Lagrangian\\Path Integral
Quantization:\\Effective Gauge Theories}
\author{Carsten Grosse-Knetter\thanks{E-Mail:
knetter@physw.uni-bielefeld.de}\\[5mm]Universit\"at Bielefeld\\
Fakult\"at f\"ur Physik\\33501 Bielefeld\\Germany}
\date{BI-TP 93/40\\hep-ph/9308201\\August 1993}
\begin{document}
\begin{titlepage}
\maketitle
\thispagestyle{empty}
\begin{abstract}
The equivalence of correct Hamiltonian and naive Lagrangian
(Faddeev--Popov) path integral quantization (Matthews's theorem)
is proven for gauge theories with arbitrary effective
interaction terms. Effective gauge-boson self-interactions and
effective interactions with scalar and fermion fields are considered.
This result becomes extended to effective gauge theories with
higher derivatives of the fields.
\end{abstract}
\end{titlepage}

\section{Introduction}
\typeout{Section 1}
\setcounter{equation}{0}
Gauge theories with arbitrary (non--Yang--Mills)
effective interaction terms have been examined in order
to parametrize possible deviations of the self interactions
of the electroweak gauge bosons \cite{nonlin,ew} and of the
gluons \cite{gluons} from the standard model predictions
with respect to experimental tests of these
couplings. Such effective Lagrangians
usually are quantized within
the Faddeev--Popov formalism \cite{fapo}, which yields
the generating functional (path integral (PI))
\begin{equation}
Z[J]=\int\Df\Phi\,\exp\left\{i\int\dx[\lag+\lag_{g.f.}+
\lag_{ghost}+J\Phi]\right\},
\label{fp}\end{equation}
where $\lag$ is the effective Lagrangian, $\lag_{g.f.}$ and
$\lag_{ghost}$ are the gauge fixing (g.f.)\
term and the ghost term which are
obtained in the standard manner. ($\Phi$ is a shorthand notation for
all fields in the quantized Lagrangian
$\lag+\lag_{g.f.}+\lag_{ghost}$.)
The generating functional (\ref{fp}) is very convenient for practical
calculations because it is manifestly
covariant (if a covariant gauge
is is chosen), it does not involve the generalized momenta of the
fields and it directly implies the Feynman rules
(i.e., the quadratic terms in $\lag+\lag_{g.f.}+\lag_{ghost}$
yield the propagators and the other terms yield the vertices
in the usual way).
However, (\ref{fp}) is derived from a naive Lagrangian PI ansatz
\cite{fapo}, while
correct quantiztion has to be
performed within the more elaborate
Hamiltonian PI formalism
\cite{fad,sen,sun,gity}. Thus,
to justify the (Lagrangian)
Faddeev--Popov PI (\ref{fp}) for effective gauge theories
one has to derive it within the Hamiltonian PI formalism, i.e.
one has to prove the
equivalence of Hamiltonian and Lagrangian PI quantization, which is
known as Matthews's theorem\footnote{Originally, the name
``Matthews's theorem'' simply denotes the statement that the Feynman
rules directly follow from the effective Lagrangian in the
usual way \cite{mat}. (Of course
for a gauge invariant Lagrangian $\lag$, the Feynman rules do not
follow from $\lag$ alone but from $\lag+\lag_{g.f.}+\lag_{ghost}$.)
Reformulated within the PI formalism, however,
this means that an arbitrary
Lagrangian can be quantized by using the naive Lagrangian PI ansatz
\cite{bedu,gk1,gk2}.}.

Matthews's theorem has been proven for Yang--Mills theories
{\em without\/} additional
effective interaction terms by Faddeev \cite{fad}
and for massive (and thus gauge
noninvarint)
Yang--Mills theories  without effective interaction
terms by Senjanovic \cite{sen}. For {\em arbitrary}
interactions of scalar
fields, this theorem has been derived by Bernard and Duncan
\cite{bedu} and for arbitrary
interactions of massive vector fields by
myself \cite{gk1}. In \cite{gk2} I have generalized these results to
effective interactions which also involve higher
derivatives of the fields.
In this article I will
complete the proof
of Matthews's theorem for arbitrary interactions of the physically
most important types of particles by considering effective
Lagrangians with massless vector fields and with fermion fields.

Massless vector fields necessarily have to be understood as gauge
fields. A Lagrangian with massless vector fields but gauge
noninvariant interactions of these would make no physical sense
because without a gauge fixing term, which only becomes
introduced for gauge invariant
Lagrangians (within the Hamiltonian PI as well
as within the Lagrangian PI), the operator occuring in the quadratic
part of the Lagrangian has no inverse and therefore it is
impossible to obtain a propagator for the vector fields.
Thus I will prove Matthews's theorem for
gauge theories with additional
arbitrary (non--Yang--Mills) self interactions of the gauge fields,
with arbitrary couplings of the gauge fields to scalar fields
and to fermion fields and with arbitrary interactions
among the scalar and fermion fields. All
effective interaction terms
are assumed to be gauge invariant. The proof also
applies to the case of spontaneously broken gauge theories (SBGTs),
i.e.\ gauge theories with massive
gauge fields, because one can assume
that the scalar fields that are coupled to the gauge fields have a
nonvanishing vacuum expectation value. Matthews's theorem
for SBGTs in which {\em all}
gauge bosons are massive has already been
derived in \cite{gk1,gk2}. There, a SBGT was rewritten as a gauge
noninvariant model by applying the Stueckelberg formalism and then
Matthews's theorem for gauge noninvariant Lagrangians was used.
In this
article I will present a more direct proof
of this theorem that does not use
the Stueckelberg formalism
and that also applies to SBGTs in which not
all gauge bosons are massive (like electroweak models).

Lagrangians with gauge fields and with fermion fields are singular.
The presence of gauge fields implies first class constraints and
the presence of fermion fields implies second class constraints.
Therefore, to prove Matthews's theorem one has to take into account
the formalism of quantization of constrained systems which goes
back to Dirac \cite{dirac} and which has been formulated
in the PI formalism by Faddeev \cite{fad} and Senjanovic \cite{sen}.
(Extensive treatises on this subject
can be found in \cite{sun,gity}.)
Within this formalism, a gauge
theory cannot be {\em directly\/} quantized
in the Lorentz-gauge or, for SBGTs,
in the $\rm R_\xi$-gauge (which are the most convenient
gauges for practical calculations) because the
corresponding g.f.\ conditions
cannot be written as relations among
the fields and the conjugate fields alone and
thus they are not g.f.\ conditions within the Hamiltonian framework.
Therefore, I will first derive the generating functional (\ref{fp})
in the Coulomb-gauge and then use the equivalence of all gauges,
i.e.\ the independence of the $S$-matrix elements from the choice of
the gauge in the Faddeev--Popov formalism \cite{lezj,able},
in order to
generalize this result to any other gauge.

To complete the proof of Matthews's theorem, one has to take into
account effective gauge theories with higher derivatives of the
fields,
which also have been investigated for phenomenological
reasons \cite{phhide}. Actually, all unphysical effects that are
connected with Lagrangians with higher derivatives
(higher-order Lagrangians) \cite{bedu,ost,hide},
namely additional degrees of freedom, unbound energy from below,
etc., are absent within the effective-Lagrangian formalism
\cite{gk2} because an effective Lagrangian
is assumed to be the low-energy approximation
of well-behaved ``new physics'', i.e.\ it parametrizes
the low-energy effects of a renormalizable theory
with heavy particles in which no higher derivatives occur.
In fact, all higher
time derivatives of the fields can be eliminated from the effective
Lagrangian by applying the
equations of motion (EOM) to the effective
interaction term (upon
neglecting higher powers of the effective coupling
constant). The (in general forbidden) use of the EOM is
correct because one can find field
transformations which have the same effect as the
application of the EOM to the effective
interaction term \cite{gk2,eom,pol}; these transformations involve
derivatives of the fields. In \cite{gk2} it has been shown
that Lagrangians which are related by such field transformations are
physically equivalent (at the classical and at the quantum level)
because these become
canonical transformations within the Hamiltonian treatment of
higher-order Lagrangians (Ostrogradsky formalism \cite{ost}).
Thus, each effective higher-order Lagrangian can be reduced to an
equivalent Lagrangian without higher time derivatives.
Since the use of the EOM does not affect the gauge invariance of a
Lagrangian, Matthews's theorem for effective gauge theories
with higher derivatives can be proven by using this reduction
and by applying Matthews's theorem for effective gauge theories
with at most first time derivatives.
Especially the treatment of fermion fields
can be simplified very much
because the EOM for these fields only depend on first
time derivatives. Therefore one can eliminate not only higher
but also first time derivatives of the fermion
fields from the effective interaction term
and thus the proof of Matthews's
theorem can be reduced
to the case of effective interactions
in which no time derivatives of
these fields occur.

In this article I will assume that
the effective interactions, which are only the
{\em deviations\/} from the standard interactions (i.e.\ from the
Yang--Mills self-interactions of the gauge fields, minimal gauge
couplings of these to the scalar and fermion fields, Yukawa
couplings and derivative-free scalar self-interctions),
are proportional to a coupling constant $\ep$ with $\ep\ll 1$.
This is justified for phenomenologically motivated effective
Lagrangians because these are studied in order to parametrize small
deviations from the standard model \cite{nonlin,ew,gluons}.
When deriving Matthews's theorem, I
will, according to \cite{bedu,gk1,gk2}, neglect higher powers
of $\ep$ and, besides, terms proportional to $\df$
which become zero if dimensional
regularization is applied.

This paper is organized as follows: In section~2 I derive the
Faddeev--Popov path integral for effective
gauge theories (without higher derivatives) by
using the Hamiltonian path integral formalism. In section~3 I
generalize this proof of Matthews's theorem to effective gauge
theories with higher derivatives by
applying the equations of motion in order to remove all higher
time derivatives from the effective interaction term.
Section~4 contains the summary of my results.

\section{Matthews's Theorem for Effective Gauge Theories}
\typeout{Section 2}
\setcounter{equation}{0}
In this section I quantize a gauge theory with an additional
arbitrary effective interaction term in the
Hamiltonian PI formalism \cite{fad,sen,sun,gity}
in order to derive the Faddeev--Popov PI (\ref{fp}).

The effective Lagrangian is given by
\begin{equation}
\lag=\lag_0+\ep\lag_I=-\frac{1}{4}F_a^{\mu\nu}F^a_{\mu\nu}
+i\bp_a\gamma^\mu D_\mu\psi_a+
(D^\mu\vp_a^\dagger)(D_\mu\vp_a)
-V(\psi_a,\bp_a,\vp_a,\vc_a)+\ep\lag_I.
\label{leff}\end{equation}
The
field strength tensor and the covariant derivatives are
\begin{eqnarray}
F_{\mu\nu}^a&\equiv&\pa_\mu A_\nu^a-\pa_\nu A_\mu^a-
gf_{abc}A_\mu^b
A_\nu^c,\label{F}\\
D_\lambda F_{\mu\nu}^a&\equiv&\pa_\lambda F_{\mu\nu}^a-gf_{abc}
A_\lambda^b F_{\mu\nu}^c,\label{DF}\\
D_\mu\psi_a&\equiv&\pa_\mu\psi_a
+igA_\mu^c t^{ab}_c\psi_b,\label{Dpsi}\\
D_\mu\bp_a&\equiv&\overline{(D_\mu\psi_a)},\\
D_\mu\vp_a&\equiv&\pa_\mu\vp_a
+igA_\mu^c \tb^{ab}_c\vp_b,\label{Dphi}\\
D_\mu\vc_a&\equiv&(D_\mu\vp_a)^\dagger.
\end{eqnarray}
(Higher covariant derivatives are defined analogously.)
$g$ is the gauge coupling constant, $f_{abc}$ are the structure
constants and $t^{ab}_c$ and $\tb^{ab}_c$ are the generators of
the gauge group  in its representation in the fermion sector
and in the
scalar sector respectively. $V(\psi_a,\bp_a,\vp_a,\vc_a)$ contains
derivative-free interactions of the fermion and scalar fields,
viz. Yukawa couplings and scalar self-interactions.

The effective interaction term $\ep\lag_I$, which parametrizes the
deviations from the minimal gauge theory, contains arbitrary
interactions of the fields which are governed by the effective
coupling constant $\ep$ with $\ep\ll 1$. As pointed out in the
introduction, an effective Lagrangian like (\ref{leff}) only has a
physical meaning if the effective interaction term is gauge
invariant. This means that {\em the gauge fields $A_\mu^a$ do not
occur arbitrarily in $\lag_I$ but only through the field strength
tensor and through covariant derivatives.\/} Furthermore,
in this section I assume that {\em $\lag_I$ does neither
depend on higher time
derivatives of the fields nor on first time
derivatives of the $A_0^a$\footnote{\rm Actually, the absence
of $\dot{A}_0^a$ already follows from the gauge invariance
and the requirement that no higher derivatives occur in $\lag_I$.}
and of the fermion fields $\psi_a$ and $\bp_a$.\/} The case of
interactions with higher derivatives will be treated in
the next section.

{}From (\ref{leff}) one finds the conjugate fields (generalized
momenta):
\begin{eqnarray}
\pi_0^a&=&\frac{\pa\lag}{\pa\dot{A}^0_a}=0,\label{p0}\\
\pi_i^a&=&\frac{\pa\lag}{\pa\dot{A}^i_a}=F_{i0}^a+\ep
\frac{\pa\lag_I}{\pa\dot{A}^i_a}=\dot{A}^i_a+\pa_iA_0^a
-gf_{abc}A_i^bA_0^c+\ep\frac{\pa\lag_I}{\pa\dot{A}^i_a}
,\label{pi}\\
\pi_\psi^a&=&\frac{\pa\lag}{\pa\dot{\psi}_a}=i\bp_a\gamma^0
,\label{ppsi}\\
\pi_\bp^a&=&\frac{\pa\lag}{\pa\dot{\bp}_a}=0,\label{ppsib}\\
\pi_\vp^a&=&\frac{\pa\lag}{\pa\dot{\vp}_a}=
D_0\vc_a+\ep\frac{\pa\lag_I}{\pa\dot{\vp}_a}=
\dot{\vp}_a^\dagger-igA_0^c\tb_c^{ba}\vc_b
+\ep\frac{\pa\lag_I}{\pa\dot{\vp}_a},\label{pphi}\\
\pi_\vci^a&=&\frac{\pa\lag}{\pa\dot{\vp}_a^\dagger}=
D_0\vp_a+\ep\frac{\pa\lag_I}{\pa\dot{\vp}_a^\dagger}=
\dot{\vp}_a+igA_0^c\tb_c^{ab}\vp_b+
\ep\frac{\pa\lag_I}{\pa\dot{\vp}_a^\dagger}.\label{pphic}
\end{eqnarray}
The relations (\ref{p0}), (\ref{ppsi}) and (\ref{ppsib}) do not
contain $\ep$-terms due to the assumption that $\lag_I$ does not
depend on $\dot{A}_0^a$, $\dot{\psi}_a$ and $\dot{\bp}_a$. These
relations cannot be solved for the velocities; they are constraints.
The remaining of the above equations can be solved for the
velocities, they become (in the first order of $\ep$):
\begin{eqnarray}
\dot{A}_a^i&=&\pi_i^a-\pa_i A_0^a+gf_{abc}A_i^bA_0^c
-\ep\frac{\pa\lag_I}{\pa\dot{A}^i_a}\Bigg|_\subs
+O(\ep^2),\label{da}\\
\dot{\vp}_a^\dagger&=&\pi_\vp^a+igA_0^c\tb^{ba}_c\vc_b
-\ep\frac{\pa\lag_I}{\pa\dot{\vp}_a}\Bigg|_\subs
+O(\ep^2),\label{dphic}\\
\dot{\vp}_a&=&\pi_\vci^a-igA_0^c\tb^{ab}_c\vp_b
-\ep\frac{\pa\lag_I}{\pa\dot{\vp}_a^\dagger}\Bigg|_\subs
+O(\ep^2).\label{dphi}
\end{eqnarray}
One obtains the Hamiltonian
\begin{eqnarray}
\ham&=&\pi_\mu^a\dot{A}_a^\mu+\pi_\psi^a
\dot{\psi}_a+\dot{\bp}_a\pi_\bp^a+
\pi_\vp^a\dot{\vp}_a+\pi_\vci^a\dot{\vp}_a
^\dagger-\lag\nonumber\\
&=&\frac{1}{2}\pi_i^a\pi_i^a-\pi_i^a\pa_i A_0^a
+gf_{abc}\pi_i^a A_i^b A_0^c+\frac{1}{4}F_{ij}^aF_{ij}^a
\nonumber\\&&
-igA_0^c t^{ab}_c(\pi_\psi^a \psi_b-\bp_a\pi_\bp^b)
+i\bp_a\gamma_iD_i\psi_a\nonumber\\
&&+\pi_\vci^a\pi_\vp^a-igA_0^c\tb_c^{ab}(\pi_\vp^a\vp_b
-\vc_a\pi_\vci^b)+(D_i\vp_a^\dagger)(D_i\vp_a)+V\nonumber\\
&&-\ep\lib(A_i^a,\psi_a,\bp_a,\vp_a,\vc_a,\pi_i^a,
\pi_\vp^a,\pi_\vci^a)+O(\ep^2)
\label{ham}\end{eqnarray}
with
\begin{equation} \lib(A_i^a,\psi_a,\bp_a,\vp_a,\vc_a,\pi_i^a,
\pi_\vp^a,\pi_\vci^a)
\equiv\lag_i\Bigg|_\subs.
\label{lib}\end{equation}

One can use the identities
\begin{eqnarray}
[D_\mu ,D_\nu ]\psi_a&=&igF^c_{\mu\nu}t^{ab}_c
\psi_b,\label{compsi}\\
{}[D_\mu, D_\nu ]\vp_a&=&igF^c_{\mu\nu}\tb^{ab}_c
\vp_b,\label{comphi}\\
{}[D_\mu,D_\nu]F^a_{\kappa\lambda}&=&-
gf_{abc}F^b_{\mu\nu}F^c_{\kappa\lambda}
\label{com}\end{eqnarray}
(and the corresponding relations for $\bp_a$ and $\vc_a$) in order
to rewrite those expressions in $\lag_I$,
in which time and spatial covariant derivatives
act on the fields,
such that the time derivatives are applied first.
Remembering the discussion of the paragraph preceding equation
(\ref{p0}) one can then easily
see that $\lag_I$ depends on the $A_0^a$ only
through the expressions
\begin{equation}
F_{i0}^a,
\qquad D_0F_{ij}^a,\qquad D_0\vp_a,\qquad D_0\vc_a
.\label{a0terms}\end{equation}
Using the relation
\begin{equation}
D_\lambda F_{\mu\nu}^a+D_\mu F_{\nu\lambda}^a
+D_\nu F_{\lambda\nu}^a=0
\label{homeq}\end{equation}
in order to to rewrite $D_0F_{ij}^a$ as
\begin{equation}
D_0F_{ij}^a=D_jF_{i0}^a-D_i F_{j0}^a
\label{d0fij}\end{equation}
and the definition (\ref{lib})
one finds that $\lib$ does not depend
on the $A_0^a$. Thus, the gauge invariance and the absence of
higher time derivatives (and of first time derivatives
of $A_0^a$, $\psi_a$ and $\bp_a$) in $\lag_I$ yields
\begin{equation}
\frac{\pa\lib}{\pa A_0^a}=0.
\label{noa0}\end{equation}

As mentioned, the relations (\ref{p0}), (\ref{ppsi}) and
(\ref{ppsib})
imply the primary constraints
\begin{eqnarray}
\phi_1^a&=&\pi_0^a=0,\label{pc}\\
\phi_\psi^a&=&\pi_\psi^a-i\bp_a\gamma^0=0,\label{pcpsi}\\
\phi_\bp^a&=&\pi_\bp^a=0.\label{pcpsib}
\end{eqnarray}
The requirement that these primary constraints have to be consistent
with the EOM, i.e.\ the demand
\begin{equation}
\dot{\phi}_a^{(1)}=\{\phi_a^{(1)},H^{(1)}\}=0,\qquad\mbox{with}
\qquad
\ham^{(1)}=\ham+\lambda_a^{(1)}\phi_a^{(1)}
\label{cons}\end{equation}
(where the $\phi^{(1)}_a$ are the primary constraints and the
$\lambda^{(1)}_a$ are Lagrange multipliers), yields secondary
constraints. Actually, (\ref{pcpsi}) and
(\ref{pcpsib}) do not imply secondary constraints, the relation
(\ref{cons}) only determines
the Lagrange multipliers corresponding to
these constraints \cite{gity}. The $\phi_1^a$ (\ref{pc})
imply the secondary constraints
\begin{equation}
\phi_2^a=\pa_i\pi_i^a+gf_{abc}\pi_i^b A_i^c
-igt_a^{bc}(\pi_\psi^b\psi_c-\bp_b\pi_\bp^c)
-ig\tb^{bc}_a(\pi_\vp^b\vp_c-\vc_b\pi_\vci^c)=0.
\label{sc}\end{equation}
Due to (\ref{noa0}),
these secondary constraints do not contain $O(\ep)$-terms,
i.e.\, they are independent of the form of
the effective interaction term $\lag_I$
(in the first order of $\ep$).
There are no tertiary, etc.\ constraints. One can easily check that
the constraints $\phi_\psi^a$ (\ref{pcpsi})
and $\phi_\bp^a$ (\ref{pcpsib}) are second class
and that
the constraints $\phi_1^a$ (\ref{pc}) and $\phi_2^a$ (\ref{sc})
are first class.

Due to the presence of the first class constraints, the solutions of
the Hamiltonian EOM contain undetermined Lagrange multipliers. To
remove these ambiguities, one has to introduce additional
gauge-fixing conditions so that constraints and g.f.\ conditions
together form a system of second class constraints which is
consistent with the EOM \cite{fad,sen,sun,gity}.
As mentioned in the introduction, the usual
Lorentz g.f.\ conditions
\begin{equation}
\chi_1^a=\pa^\mu A^a_\mu-C^a=0
\label{lg}\end{equation}
(and also the $\rm R_\xi$-g.f. conditions for SBGTs) are not g.f.
conditions within the Hamiltonian formalism
\cite{fad,sen,sun,gity}
because they are not
relations among the fields and the conjugate fields alone due to the
presence of the velocities $\dot{A}_0^a$ in (\ref{lg}),
which cannot be expressed in
terms of the momenta. Therefore I quantize the effective gauge theory
within the Coulomb-gauge, i.e.\ by choosing the primary g.f. conditions
\begin{equation}
\chi_1^a=\pa^i A^a_i-C^a=0.
\label{pgf}\end{equation}
(Instead of the
Coulomb--gauge, one can alternatively choose the axial gauge or, for
SBGTs, the unitary gauge \cite{gk1}).
Next, one has to construct secondary g.f.\ conditions
$\chi_2^a$ by demanding
\begin{equation}
\{\chi_1^a,H\}=0
\end{equation}
which ensures the consitency with the EOM \cite{sun,gity}.
One finds\footnote{The g.f.\ conditions (\ref{pgf})
and (\ref{sgf}) do not
fulfil the condition $\{\chi_1^a,\chi_2^b\}=0$ required in
\cite{fad,sen}. However, this demand is unnecessary
\cite{sun,gity,kadi}.}
\begin{equation}
\chi_2^a=\Delta A_0^a-\pa_i\pi_i^a
-gf_{abc}\pa_i(A_i^bA_0^c)
+\ep\pa_i\frac{\pa\lib}{\pa\pi_i^a}=0.
\label{sgf}\end{equation}

The Hamiltonian path intergral
\cite{fad,sen,sun,gity} for this system
is given by\footnote{For convenience, I introduce the source terms
in the PI after all manipulations have been done. (The source terms
for the ghost fields have to be introduced later, anyway.) Actually,
if the source terms would be considered from the beginning, the
subsequent procedure would not leave them unchanged. However, a
change in the source terms does not effect the $S$-matrix elements
\cite{able}.}
\begin{eqnarray}
Z& =&\int\Df A_\mu^a\Df\psi_a
\Df\bp_a\Df \vp_a\Df\vc_a
\Df \pi_\mu^a\Df\pi_\psi^a\Df\pi_\bp^a\Df\pi_\vp^a\Df\pi_\vci^a\,
\nonumber\\&&\times \exp\left\{i\int\dx\left[\pi_\mu^a
\dot{A}_a^\mu+\pi_\psi^a\dot{\psi}_a+\dot{\bp}_a\pi_\bp^a+
\pi_\vp^a\dot{\vp}_a+\pi_\vci^a\dot{\vc}_a-\ham\right]\right\}
\nonumber\\&&
\times\delta(\phi_\psi^a)\delta(\phi_\bp^a)\delta(\phi_1^a)
\delta(\phi_2^a)\delta(\chi_1^a)\delta(\chi_2^a)\nonumber\\&&
\times{\rm Det^{\frac{1}{2}}}\,(\{\Phi_{2nd}^a(\vx),
\Phi_{2nd}^b(\vy)\}\dzero)
\Det(\{\Phi_{1st}^a(\vx), \X^b(\vy)\}\dzero),
\label{hpi}\end{eqnarray}
where $\Phi^a_{2nd}$, $\Phi^a_{1st}$ and $\X^a$ denote
{\em all\/} second class constraints, first class constraints
and g.f.\ conditions respectively.
First let me consider the determinants\footnote{The
factors $\delta(x^0-y^0)$ in the
arguments of the
determinants are missing in \cite{fad,sen,sun,gity}.
However, they
neccesarily have to be present because ${\rm Det}\,
\{\Phi^a_{1st}({\bf x}),\X^b({\bf y})\}$
(where
$\Phi^a_{1st}({\bf x})$ and $\X^b({\bf y})$ are taken at equal times)
has to be introduced for {\em all\/} times and ${\rm Det}\,(
\{\Phi^a_{1st}({\bf x}),\X^b({\bf y})\}\delta(x^0-y^0))$
is the ``product'' of this expression over
all times. (The same is true for the other determinant in
(\ref{hpi}).)}
occuring in (\ref{hpi}). The fundamental Poisson brackets
immediately imply
\begin{equation}
\mbox{Det}^{\frac{1}{2}}\,
(\{\Phi_{2nd}^a(\vx),\Phi_{2nd}^b(\vy)\}\dzero)
=\mbox{constant}.
\end{equation}
Therefore, this term can be neglected in the PI. Furthermore, one
finds
\begin{eqnarray}
\{\phi_1^a(\vx),\chi_1^b(\vy)\}
&=&0,\\ \!\!\!\!\!\!\!\!\! - \{\phi_1^a(\vx),\chi_2^b(\vy)\}
= \{\phi_2^a(\vx),\chi_1^b(\vy)\}
&=&(\delta_{ab}\Delta+gf_{abc}(\pa_i A_i^c)+
gf_{abc}A_i^c\pa_i)\delta^3(\vx-\vy).\label{det21}
\label{const}\end{eqnarray}
The absence of $O(\ep)$-terms in (\ref{det21})
is again a consequence of (\ref{noa0}). This yields
\begin{equation}
\Det(\{\Phi_{1st}^a(\vx),\X^b(\vy)\}\dzero)
=\,\mbox{Det}^2\,[(\delta_{ab}\Delta+gf_{abc}(\pa_i A_i^c)+
gf_{abc}A_i^c\pa_i)\delta^4(x-y)].
\label{ghostdet}\end{equation}

The following steps
are very simlar to those made either within the
Hamiltonian PI quantization
of a Yang--Mills theory without effective interaction
terms in \cite{sun,gity} or within the treatment of gauge
noninvariant
effective Lagrangians in \cite{bedu,gk1}. Therefore, I will discuss
them only very briefly. First one observes that $\ham$ (\ref{ham})
contains a term $A_0^a\phi_2^a$. Due to the presence of
$\delta(\phi_2^a)$ in the PI
this term can be ommited. Then one integrates over
$\pi_0^a$, $\pi_\psi^a$ and $\pi_\bp^a$ and finds
\begin{eqnarray}
Z& =&\int\Df A_\mu^a\Df\psi_a
\Df\bp_a\Df \vp_a\Df\vc_a
\Df \pi_i^a\Df\pi_\vp^a\Df\pi_\vci^a\,\nonumber\\&&
\times\exp\Bigg\{i\int\dx\Bigg[-\frac{1}{2}\pi_i^a\pi_i^a
+\pi_i^a\dot{A}^i_a-\frac{1}{4}F_{ij}^aF_{ij}^a\nonumber\\&&\quad
+i\bp_a\gamma^0\dot{\psi}_a-i\psi_a\gamma_i D_i\psi_a
-\pi_\vp^a\pi_\vci^a+\pi_\vp^a\dot{\vp}_a+\pi_\vci^a\dot{\vp}_a
^\dagger-(D_i\vp_a^\dagger)(D_i\vp_a)-V\nonumber\\&&\quad+\ep
\lib(A_i^a,\psi_a,\bp_a,\vp_a,\vc_a,\pi_i^a,
\pi_\vp^a,\pi_\vci^a)\Bigg]\Bigg\}\nonumber\\&&\times
\delta(\tilde{\phi}_2^a)\delta(\chi_1^a)
\delta(\chi_2^a)
\Det(\{\Phi_{1st}^a(\vx),\X^b(\vy)\}\dzero)
\end{eqnarray}
with
\begin{equation}
\tilde{\phi}_2^a=\pa_i\pi_i^a+gf_{abc}\pi_i^b A_i^c +
g\bp^b\gamma^0t_a^{bc}\psi_c-ig\tb_a^{bc}
(\pi_\vp^b\vp_c-\vc_b\pi_\vci^c)=0.
\label{sct}\end{equation}
After rewriting
\begin{equation}
\delta(\chi_2^a)=\delta(A_0^a-\tilde{A}_0^a)\,\mbox{Det}^{-1}
[(\delta_{ab}\Delta+gf_{abc}(\pa_i A_i^c)+
gf_{abc}A_i^c\pa_i)\delta^4(x-y)]
\label{deti}\end{equation}
(where $\tilde{A}_0^a$ is the solution of the differential
equation (\ref{sgf}) with the boundary condition
that $\tilde{A}_0^a$ vanishes for $|\vx|\to\infty$)
one can also integrate over $A_0^a$.
Due to (\ref{noa0}), the argument of the
determinant in (\ref{deti}) also
does not contain $O(\ep)$-terms and, besides, the integration over
$A_0^a$ does not affect $\lib$.
Next one reintroduces the variables $A_0^a$ by using
\begin{equation}
\delta(\tilde{\phi}_2^a)=
\int\Df A_0^a\,\exp\left\{-i\int\dx A_0^a\tilde{\phi}_2^a
\right\}
\end{equation}
and gets
\begin{eqnarray}
Z& =&\int\Df A_\mu^a\Df\psi_a
\Df\bp_a\Df \vp_a\Df\vc_a
\Df \pi_i^a\Df\pi_\vp^a\Df\pi_\vci^a\,\nonumber\\&&
\times\exp\Bigg\{i\int\dx\Bigg[-\frac{1}{2}\pi_i^a\pi_i^a
+\pi_i^a F_{i0}^a
-\frac{1}{4}F_{ij}^aF_{ij}^a\nonumber\\&&\quad
+i\psi_a\gamma^\mu D_\mu\psi_a
-\pi_\vp^a\pi_\vci^a+\pi_\vp^a D_0\vp_a
+\pi_\vci^a D_0\vc_a
-(D_i\vp_a^\dagger)(D_i\vp_a)
-V\nonumber\\&&\quad+\ep
\lib(A_i^a,\psi_a,\bp_a,\vp_a,\vc_a,\pi_i^a,
\pi_\vp^a,\pi_\vci^a)\Bigg]\Bigg\}\nonumber\\&&\times
\delta(\pa^iA_i^a-C^a)
\Det[(\delta_{ab}\Delta+gf_{abc}(\pa_i A_i^c)+
gf_{abc}A_i^c\pa_i)\delta^4(x-y)].
\end{eqnarray}
In order to obtain expressions quadratic in
the momenta, one rewrites
this as
\begin{eqnarray}
Z& =&\int\Df A_\mu^a\Df\psi_a
\Df\bp_a\Df \vp_a\Df\vc_a\nonumber\\&&
\times
\exp\left\{i\ep\int\dx\lib\left(A_i^a,\psi_a,\bp_a,\vp_a,\vc_a,
\frac{\delta}{i\delta K_i^a},\frac{\delta}{i\delta K_\vp^a},
\frac{\delta}{i\delta K_\vci^a}\right)\right\}\nonumber\\&&
\times\int\Df\pi_i^a\Df\pi_\vp^a\Df\pi_\vci^a
\exp\Bigg\{i\int\dx\Bigg[-\frac{1}{2}\pi_i^a\pi_i^a
+\pi_i^a F_{i0}^a
-\frac{1}{4}F_{ij}^aF_{ij}^a\nonumber\\&&\quad
+i\psi_a\gamma^\mu D_\mu\psi_a
-\pi_\vp^a\pi_\vci^a+\pi_\vp^a D_0\vp_a
+\pi_\vci^a D_0\vc_a
-(D_i\vp_a^\dagger)(D_i\vp_a)
-V\nonumber\\&&\quad +K_i^a\pi_i^a+K_\vp^a\pi_\vp^a
+K_\vci^a\pi_\vci^a\Bigg]\Bigg\}
\Bigg|_{\S K_i^a=K_\vp^a=K_\vci^a=0}\nonumber\\&&\times
\delta(\pa^iA_i^a-C^a)
\Det[(\delta_{ab}\Delta+gf_{abc}(\pa_i A_i^c)+
gf_{abc}A_i^c\pa_i)\delta^4(x-y)].
\end{eqnarray}
Now one can
do the Gaussian integrations over the momenta. With
$\lag_0$ given in (\ref{leff}) one finds
\begin{eqnarray}
Z& \!=\!&\int\Df A_\mu^a\Df\psi_a
\Df\bp_a\Df \vp_a\Df\vc_a\,
\exp\left\{i\int\dx\lag_0\right\}
\nonumber\\&&\times
\exp\left\{i\ep\int\dx\lib\left(A_i^a,\psi_a,\bp_a,\vp_a,\vc_a,
\frac{\delta}{i\delta K_i^a},\frac{\delta}{i\delta K_\vp^a},
\frac{\delta}{i\delta K_\vci^a}\right)\right\}\nonumber\\&&
\times\exp\left\{i\int\dx\left[\frac{1}{2}K_i^aK_i^a
+K_\vp^a K_\vci^a+K_i^a F_{i0}^a+K_\vp^aD_0\vc_a+
K_\vci^aD_0\vp_a\right]\right\}
\Bigg|_{\S K_i^a=K_\vp^a=K_\vci^a=0}\nonumber\\&&
\times\delta(\pa^iA_i^a-C^a)
\Det[(\delta_{ab}\Delta+gf_{abc}(\pa_i A_i^c)+
gf_{abc}A_i^c\pa_i)\delta^4(x-y)].
\end{eqnarray}
This expression can be simplified in complete analogy to the
procedure outlined in \cite{bedu,gk1}.
Thus I only present the result which is found
after some calculations (by neglecting $O(\ep^2)$ and
$\df$ terms), viz.
\begin{eqnarray}
Z& =&\int\Df A_\mu^a\Df\psi_a
\Df\bp_a\Df \vp_a\Df\vc_a\,
\exp\left\{i\int\dx[\lag_0+\ep\tilde{\lag}_I]\right\}
\nonumber\\&&\times
\delta(\pa^iA_i^a-C^a)
\Det[(\delta_{ab}\Delta+gf_{abc}(\pa_i A_i^c)+
gf_{abc}A_i^c\pa_i)\delta^4(x-y)],
\label{fpcg}\end{eqnarray}
where $\tilde{\lag}_I$ turns out to be
\begin{equation}
\tilde{\lag}_I=\lib\Bigg|_\ssubs=\lag_I.
\label{a}\end{equation}
(\ref{fpcg}) with (\ref{a})
is identical to the result obtained in
the Faddeev--Popov formalism
by choosing the (Coulomb) g.f.\ conditions $\chi_1^a$
(\ref{pgf}) because the change of $\chi_1^a$ under infinitesimal
variations of the gauge parameter $\alpha_b$ is
\begin{equation}
\frac{\delta\chi_1^a(x)}{\delta\alpha_b(y)}=
(\delta_{ab}\Delta+gf_{abc}(\pa_i A_i^c)+
gf_{abc}A_i^c\pa_i)\delta^4(x-y).
\end{equation}
To derive the form (\ref{fp}) of the generating
functional one has, as usual,  to
construct
the g.f.\ term by using the $\delta$-fuction and to rewrite
the determinant as a ghost term.
Finally the source terms have to be added.
It is essential for the derivation of this result
that, due to (\ref{noa0}), no $O(\ep)$-terms occur in
the argument of the determinant in (\ref{fpcg}).
Thus the ghost term is independent of the form of the effective
interaction term as in the Faddeev--Popov formalism.
Due to the equivalence of all gauges \cite{lezj,able}
the result (\ref{fp}) can be rewritten in any other gauge
which can be derived within the Faddeev--Popov formalism,
e.g. in the
Lorentz-gauge or in the $\rm R_\xi$-gauge (for SBGTs).

The gauge
theory given by (\ref{leff}) is spontaneously broken if the vacuum
expectation value of the scalar fields (implied by the scalar
self-interactions in $V$) is nonzero; this does not affect the above
proof. Actually, this proof holds for both,
SBGTs with a {\em linearly\/}
realized scalar sector, which contain (a) physical Higgs boson(s)
\cite{ew}, and gauged nonlinear $\sigma$-models, i.e.\
SBGTs with a {\em nonlinearly\/} realized
scalar scalar sector and without
physical Higgs bosons \cite{nonlin}, because the latter
ones can be obtained from the first ones by making a point
transformation (which does not affect the Hamiltonian PI
\cite{fad,sen,sun,gity}) in order to rewrite the scalar sector
nonlinearly \cite{gk1,gk2,lezj,clt,gkko} and then taking the
limit $M_H\to \infty$ \cite{nonlin,gk1,gk2,gkko}.
Thus, for an arbitrary effective gauge theory (without
higher derivatives) the simple
Faddeev--Popov PI can be derived within the
correct Hamiltonian PI formalism.

\section{Effective Gauge Theories with Higher Derivatives}
\typeout{Section 3}
\setcounter{equation}{0}
In this section I generalize the results of the preceding one to
effective gauge theories with higher time derivatives.

Each effective Lagrangian like (\ref{leff}) can be reduced
to a Lagrangian without higher time derivatives (and also without
first time derivatives of the $A_0^a$, $\psi_a$ and $\bp_a$)
because {\em the equations of motion
following from $\lag_0$ in (\ref{leff}) can be
applied
in order to convert the effective interaction term $\lag_I$\/}
\cite{gk2} (upon neglecting higher powers of $\ep$). This statement
is nontrivial because, in general, the EOM must not be inserted into
the Lagrangian. However, one can find field transformations which
have the same effect as the application of the EOM to the effective
interaction term \cite{gk2,eom,pol}.
Actually, a field
transformation
\begin{equation}
\Phi\to\Phi+\ep T
\label{trafo}\end{equation}
(where $\Phi$ may represent any field occuring in $\lag$ and $T$ is
an arbitrary function of the fields and their derivatives)
applied to
(\ref{leff}) yields an extra term
\begin{equation}
\ep T\left(\frac{\pa\lag_0}{\pa\Phi}-\pa^\mu\frac {\pa\lag_0}
{\pa(\pa^\mu\Phi)}\right)+O(\ep^2)
\label{extra}\end{equation}
to the effective Lagrangian. Lagrangians
that are related by field
transformations like (\ref{trafo}) are physically equivalent (at the
classical and at the quantum level) \cite{gk2}
although these transformations involve derivatives
of the fields (contained in $T$) because they
become point transformations (and thus canonical
transformations) within the Hamiltonian formalism for Lagrangians
with higher derivatives (Ostrogradsky formalism \cite{ost}).
The reason for this is that in the Ostrogradsky formalism for an
$N$-th order Lagrangian all derivatives of the fields up to the
order $N-1$ are treated as
independent coordinates, and the order $N$
can be chosen
arbitrarily high without affecting the physical content
of the theory \cite{gity,gk2}.

An arbitrary effective gauge theory can be reduced
to one of the type considered in
the previous section as follows: Due to the gauge invariance,
derivatives of the fields occur in the effective interaction term
only as covariant derivatives or through the field strength tensor.
Using the identities (\ref{compsi}), (\ref{comphi}) and (\ref{com})
(and the corresponding relations
for $\bp_a$ and $\vc_a$) one again
rewrites all expressions in $\lag_I$
such that the covariant time derivatives are applied
to the fields before the
covariant spatial derivatives. Then,
higher time derivatives (and first time derivatives of
$A_0^a$, $\psi_a$ and $\bp_a$) occur in $\lag_I$ only
through the expressions
\begin{equation}
D_0 F^a_{0i}, \qquad D_0D_0 F_{ij}^a,\qquad
D_0\psi_a,\qquad D_0\bp_a,\qquad
D_0D_0\vp_a,\qquad D_0D_0\vc_a
\label{hdterms}\end{equation}
and even higher derivatives of these terms.
After using (\ref{com}) and (\ref{homeq}) in order to rewrite
$D_0D_0 F_{ij}^a$ as
\begin{equation}
D_0D_0 F_{ij}^a=D_iD_0 F_{0j}^a-D_jD_0 F_{0i}^a
-2gf_{abc}F_{i0}^b F_{j0}^c
\end{equation}
one can convert the terms (\ref{hdterms}) to terms
without higher time derivatives (and without first time
derivatives of $A_0^a$, $\psi_a$ and $\bp_a$) by using the EOM
following from $\lag_0$, viz.\
\begin{eqnarray}
D_0F_{0i}^a&=&D_jF^a_{ji}+g\bp_b\gamma_it^{bc}_a\psi_c
-ig\tb^{bc}_a\left((D_i\vp_b^\dagger)\vp_c
-\vc_b(D_i\vp_c)\right), \label{EOMA}\\
D_0\psi_a&=&\gamma^0\left(\gamma_iD_i\psi_a
-i\frac{\pa V}{\pa\bp_a}\right),\label{EOMpsi}\\
D_0D_0\vp_a&=&D_iD_i\vp_a
-\frac{\pa V}{\pa\vc_a}\label{EOMphi}
\end{eqnarray}
(and the corresponding equations for $\bp_a$ and $\vc_a$).
By repeated application of the EOM one can eliminate all
higher time derivatives from $\lag_I$.
The fact that the EOM do not contain second time derivatives of
$A_0^a$, $\psi_a$ and $\bp_a$ makes it possible to
eliminate not only higher but also
first time derivatives of these fields.
The Lagrangian obtained by applying the EOM is gauge
invariant, too,
because the form of the EOM is invariant under gauge
transformations.

Now Matthews's theorem for effective gauge theories with higher
time derivatives can be
proven as follows:
\begin{enumerate}
\item Given an arbitrary gauge invariant
effective Lagrangian $\lag$, this can be reduced to
an equivalent gauge invariant
Lagrangian $\lag_{red}$ without higher time
derivatives (and without
first time derivatives of $A_0^a$, $\psi_a$
and $\bp_a$)
by applying the EOM, i.e.\  actually by making field
transformations like (\ref{trafo}). This does not affect the
Hamiltonian PI \cite{gk2}.
\item $\lag_{red}$ can be quantized within the Hamiltonian PI
formalism by applying Matthews's theorem for first-order Lagrangians
derived in section~2. This yields the PI
\begin{equation}
Z=\int\Df\Phi\,\exp\left\{i\int\dx[\lag_{red}+\lag_{g.f.}+
\lag_{ghost}]\right\}.
\label{lred}\end{equation}
\item Going reversely through the Faddeev--Popov
procedure one can
rewrite (\ref{lred}) (after
introducing an infinite constant into the PI) as
\begin{equation}
Z=\int\Df\Phi\,\exp\left\{i\int\dx\lag_{red}
\right\}.\label{lred2}
\end{equation}
\item Within the Lagrangian PI (\ref{lred2})
the field transformations (\ref{trafo})
applied in step~1 are done inversely in order to
reconstruct the primordial effective
Lagrangian\footnote{Note that the use of the
transformations (\ref{trafo}) in (\ref{lred}) would result in an
application of the EOM following from $\lag_0+\lag_{g.f.}
+\lag_{ghost}$ (and not from $\lag_0$ alone)
which would not yield the desired result.}.
One obtains
\begin{equation} Z=\int\Df\Phi\,\exp\left\{i\int\dx\lag
\right\}.\label{lpi}
\end{equation}
The Jacobian determinant implied by
change of the functional integration measure
corresponding to
these transformations only yields extra $\df$ terms \cite{pol}
which are neglected here.
\item Applying the Faddeev--Popov formalism\footnote{In
distiction from naive
Lagrangian PI quantization, (\ref{lpi}) is not taken as an ansatz
here but
it has been derived from the Hamiltonian PI.}
to (\ref{lpi}) and adding the source terms
one finally finds (\ref{fp}) in an arbitrary gauge.
\end{enumerate}
This completes the proof of Matthews's theorem for effective gauge
theories.

\section{Summary}
\typeout{Section 4}
\setcounter{equation}{0}
In this article I have completed the proof of Matthews's theorem for
arbitrary interactions of the physically most important types of
particles. I have shown, that a gauge theory
with an arbitrary effective interaction term can
be quantized by using the convenient (Lagrangian)
Faddeev--Popov path integral because this can
be derived from the correct Hamiltonian path integral.
Thus Hamiltonian and Lagrangian path integral
quantization are equivalent. This means that the Feynman rules
can be obtained in the usual way from the effective Lagrangian.

Matthews's theorem also applies to effective gauge theories with
higher derivatives of the fields. Each effective
gauge theory can be reduced to a
gauge theory with at most first time
derivatives by applying the equations of motion to the effective
interaction term. Thus, an effective higher-order Lagrangian can
formally be treated in the same way as
a first-order one; all unphysical effects,
which normally occur when dealing with higher-order Lagrangains, are
absent because an {effective}
Lagrangian is assumed to parametrize the
low-energy effects of well-behaved ``new physics''.

Actually, these results justify the straightforward
treatment of effective gauge theories in the phenomenological
literature.

\section*{Acknowledgement}
I am very grateful to
Reinhart K\"ogerler for many interesting and important discussions.
I thank Mikhail Bilenky, Ingolf Kuss and Dieter
Schildknecht for useful
conversations about the phenomenological aspects
of effective gauge theories.

\end{document}